\documentclass[conference, twocolumn, 10pt]{IEEEtran}

\usepackage{ifpdf}
\usepackage{comment}
\usepackage{textcomp}
\usepackage{stfloats}
\usepackage{adjustbox}

\usepackage{epsfig}
\usepackage{graphics}
\usepackage{caption}
\usepackage[font=footnotesize]{subfig}

\usepackage{amsmath}
\usepackage{amsfonts}
\usepackage{amssymb}  
\usepackage{amsthm}
\usepackage{mathtools}
\usepackage{bbm}
\usepackage{empheq}
\usepackage{cases}
\usepackage[acronym]{glossaries}
\usepackage{algorithm}
\usepackage{algorithmicx}
\usepackage{algpseudocode}

\algnewcommand\algorithmicinput{\textbf{Input:}}
\algnewcommand\INPUT{\item[\algorithmicinput]}
\algrenewcommand\algorithmicrequire{\textbf{Initialize:}}
\algnewcommand\INIT{\item[\algorithmicrequire]}
\algblockdefx{MRepeat}{EndRepeat}{\textbf{repeat}}{}
\algnotext{EndRepeat}
\algrenewcommand\algorithmicforall{\textbf{for each}}
\algnewcommand{\algorithmicgoto}{\textbf{go to}}%
\algnewcommand{\Goto}[1]{\algorithmicgoto~Step \ref{#1}}%

\usepackage{array}

\usepackage{cite}

\usepackage{url}
\usepackage{hyperref}

\usepackage{xcolor}
\usepackage{enumitem}

\IEEEoverridecommandlockouts

\theoremstyle{definition}

\DeclareMathAlphabet\mathcallite{OMS}{cmsy}{m}{n}
\DeclareMathAlphabet\mathbfcal{OMS}{cmsy}{b}{n}

\DeclareMathOperator*{\minimize}{minimize}
\DeclareMathOperator*{\maximize}{maximize}

\newacronym{NES}{NES}{network energy savings}
\newacronym{LA}{LA}{link adaptation}
\newacronym{gNB}{gNB}{gNodeB}
\newacronym[plural=user equipment, firstplural=user equipment (UEs)]{UE}{UE}{user equipment}
\newacronym{MCS}{MCS}{modulation and coding scheme}
\newacronym{3GPP}{3GPP}{third generation partnership project}
\newacronym{SU-MIMO}{SU-MIMO}{single user MIMO}
\newacronym{MIMO}{MIMO}{multiple input multiple output}
\newacronym{MAC}{MAC}{medium access control}
\newacronym{5G}{5G}{fifth generation}
\newacronym{5GA}{5G-A}{fifth generation-advanced}
\newacronym{6G}{6G}{sixth generation}
\newacronym{SM}{SM}{spatial mode}
\newacronym{RB}{RB}{resource block}
\newacronym{DL}{DL}{downlink}
\newacronym{UL}{UL}{uplink}
\newacronym{CSI-RS}{CSI-RS}{channel state information reference signal}
\newacronym{CSI}{CSI}{channel state information}
\newacronym{BLER}{BLER}{block error rate}
\newacronym{PSD}{PSD}{power spectral density}
\newacronym{SINR}{SINR}{signal-to-interference-plus-noise ratio}
\newacronym{TTI}{TTI}{transmission time interval}
\newacronym{PF}{PF}{proportional fair}
\newacronym{TRX}{TRX}{transceiver chain}
\newacronym{OFDM}{OFDM}{orthogonal frequency division multiplexing}
\newacronym{OFDMA}{OFDMA}{orthogonal frequency division multiple access}
\newacronym{BLA}{BLA}{baseline LA}
\newacronym{CQI}{CQI}{channel quality indicator}
\newacronym{RI}{RI}{rank indicator}
\newacronym{PMI}{PMI}{precoder matrix indicator}
\newacronym{MIESM}{MIESM}{mutual information effective signal-to-noise ratio mapping}
\newacronym{POLITE}{POLITE}{power optimization for low interference and throughput enhancement}
\newacronym{FIFO}{FIFO}{first-in first-out}
\newacronym{PHY}{PHY}{physical}
\newacronym{PC}{PC}{power consumption}
\newacronym{KPI}{KPI}{key performance indicator}
\newacronym{UPT}{UPT}{user perceived throughput}
\newacronym{NEC}{NEC}{network energy consumption}
\newacronym{PA}{PA}{power amplifier}
\newacronym{RF}{RF}{radio frequency}
\newacronym{CSP}{CSP}{communication service provider}
\newacronym{mMIMO}{mMIMO}{massive multiple-input multiple-output}
\newacronym{EE}{EE}{energy efficiency}
\newacronym{KVI}{KVI}{key value indicator}
\newacronym{KV}{KV}{key value}
\newacronym{RAN}{RAN}{radio access network}
\newacronym{eMIMO}{XL MIMO}{extremely large MIMO}
\newacronym{QoS}{QoS}{quality of service}
\newacronym{RRM}{RRM}{radio resource management}
\newacronym{IPV}{IPV}{interference power value}
\newacronym{FTP3}{FTP3}{File Transfer Protocol 3}

\title{Energy Efficient Downlink mMIMO Using Dynamic Antenna and Power Adaptation}

\author{\IEEEauthorblockN{Ravi Sharan B A G\IEEEauthorrefmark{1}, Maliha Jada\thanks{The research work presented in this publication partially contributes to Maliha's doctoral thesis requirements at Aalto University, Finland.} \IEEEauthorrefmark{2}, Anders Karstensen\IEEEauthorrefmark{3}, Daniela Laselva\IEEEauthorrefmark{3}, Jyri H{\"a}m{\"a}l{\"a}inen\IEEEauthorrefmark{4}, Silvio Mandelli\IEEEauthorrefmark{1}}
\IEEEauthorblockA{\IEEEauthorrefmark{1} Nokia Bell Labs, Stuttgart, Germany, Email: \{ravi.bhagavathula, silvio.mandelli\}@nokia-bell-labs.com}
\IEEEauthorblockA{\IEEEauthorrefmark{2} Nokia, Finland, Email: maliha.jada@nokia.com}
\IEEEauthorblockA{\IEEEauthorrefmark{3} Nokia, Denmark, Email: \{anders.karstensen, daniela.laselva\}@nokia.com}
\IEEEauthorblockA{\IEEEauthorrefmark{4} Aalto University, Finland, Email: jyri.hamalainen@aalto.fi}
}

\begin{document}

\maketitle
\IEEEoverridecommandlockouts
\IEEEpubid{\begin{minipage}{\textwidth}\ \\[12pt] \centering
\textbf{This work has been submitted to the IEEE for possible publication. Copyright may be transferred without notice, after which this version may no longer be accessible.}
\end{minipage}}

\begin{abstract}
Massive multiple-input multiple-output (mMIMO) technology and its future evolutions are expected to address the high data rate demands of sixth generation (6G) communication systems. At the same time, network energy savings (NES) is essential in reducing the operational costs and meeting the sustainability goals of network operators. In this regard, we propose a dynamic scheme for joint antenna and power adaptation to improve NES from a user scheduling and resource allocation perspective. Antenna adaptation is performed using the multiple channel state information resource signal (CSI-RS) framework. Furthermore, the recently introduced transmit power-aware link adaptation scheme, referred to as POLITE for short, is used as the power adaptation technique. The proposed scheme adapts to variations in users' instantaneous traffic and channel conditions to opportunistically maximize NES while also inherently accounting for the user throughput. Numerical simulation results show that the proposed scheme consistently achieves a balance between NES and user perceived throughput (UPT) for different network load conditions. Especially in low and light load conditions, the proposed scheme significantly improves the intra-cell interference and boosts the overall NES, while ensuring that UPT is unaffected.
\end{abstract}


\section{Introduction}
\label{sec:introduction}

\IEEEpubidadjcol
Standardization efforts are currently ongoing for the future \gls{5GA} and \gls{6G} cellular wireless communication systems. These systems are envisioned to drive the advancements in emerging applications such as collaborative robots, digital twins and extended reality applications~\cite{muusitalo2025towards6g}. \Gls{mMIMO} is expected to play an instrumental role in addressing the high data rate demands of both \gls{5GA} and \gls{6G} systems~\cite{zwang2024tutorialemimo}. This is due to the array gain and spatial multiplexing capabilities offered by large antenna arrays in \gls{mMIMO}~\cite{bjornson2017massive}. At the same time, it is becoming increasingly important for network operators to offer wireless services with reduced energy consumption and overall carbon footprint. However, maintaining a balance between \gls{NES} and the throughput gains can be extremely challenging. This is because next generation \gls{mMIMO} \glspl{gNB} are expected to use even larger antenna arrays with $256$ antenna ports and array sizes exceeding $1000$ antenna elements~\cite{swesemann2023eemimo}.

Recent studies in~\cite{3gpp2023studynes,dlperez2022survey5gee,swesemann2023eemimo,dlaselva2024pathto6G} present several techniques to reduce \gls{PC} in next generation \gls{mMIMO} \glspl{gNB}. Of these, antenna and power adaptation have been identified as two important \gls{NES} techniques in \gls{DL}. Antenna adaptation is a spatial domain technique, where \gls{gNB} operates with only a subset of the available antenna ports. In this regard,~\cite{3gpp2023studynes} introduced a framework for antenna adaptation using multiple \gls{CSI-RS} configurations. Here, each \gls{CSI-RS} configuration is associated with a predefined number of antenna ports at \gls{gNB}, thus enabling antenna adaptation using the \acrshort{CSI} of associated \gls{UE}. In contrast, power adaptation is a power domain technique which focuses on the dynamic optimization of transmit \gls{PSD} in \gls{DL}. \Gls{POLITE}~\cite{smandelli2021polite} is one such recent scheme, where efficient \gls{LA} is used to adaptively scale down \gls{PSD}. Especially, \gls{POLITE} efficiently utilizes the time-frequency resources to make \gls{DL} transmissions robust to \gls{PSD} reduction. Extending these approaches, in this work we propose a \textit{dynamic scheme for joint antenna and power adaptation}, which aims to opportunistically improve \gls{NES} while also inherently accounting for the user throughput. 

\IEEEpubidadjcol
In the existing literature on \gls{NES}, antenna and power adaptation schemes are treated separately. The work in~\cite{mmahossain2018eemmimoload} proposed an antenna selection scheme based on the long-term temporal variations in the number of active \glspl{UE}. A deep learning-based antenna selection scheme is proposed in~\cite{nrajapaksha2024minimizing} to minimize \gls{PC} subject to per-\gls{UE} minimum rate constraints. Here, a neural network is trained to imitate the behavior of a search-based greedy heuristic to essentially determine the set of active antennas based on the \acrshort{CSI} of the associated \glspl{UE}. In~\cite{wparmudito2020loadawareee}, the authors proposed a scheme to switch-off antenna elements for prolonged durations based on \glspl{UE}' \acrshort{CSI}. The recent work in~\cite{peschiera2025optimizingtimespacepowerdomain} considered a \gls{PC} minimization problem spanning time, space and power domains. Here, transmit power allocation is performed per each active antenna based on per-\gls{UE} minimum rate requirements. However, this does not translate into \gls{PSD} reduction, which involves efficient resource allocation based on both \gls{UE} traffic and the \acrshort{CSI}. In contrast to the prior work, in this work the interplay between \gls{LA} and resource allocation aspects of antenna and power adaptation schemes is leveraged to efficiently balance the user throughput and \gls{NES}. 

\section{Network Model}
\label{sec:systemmodel}

\begin{figure}
    \centering
 \includegraphics[width=0.8\linewidth]{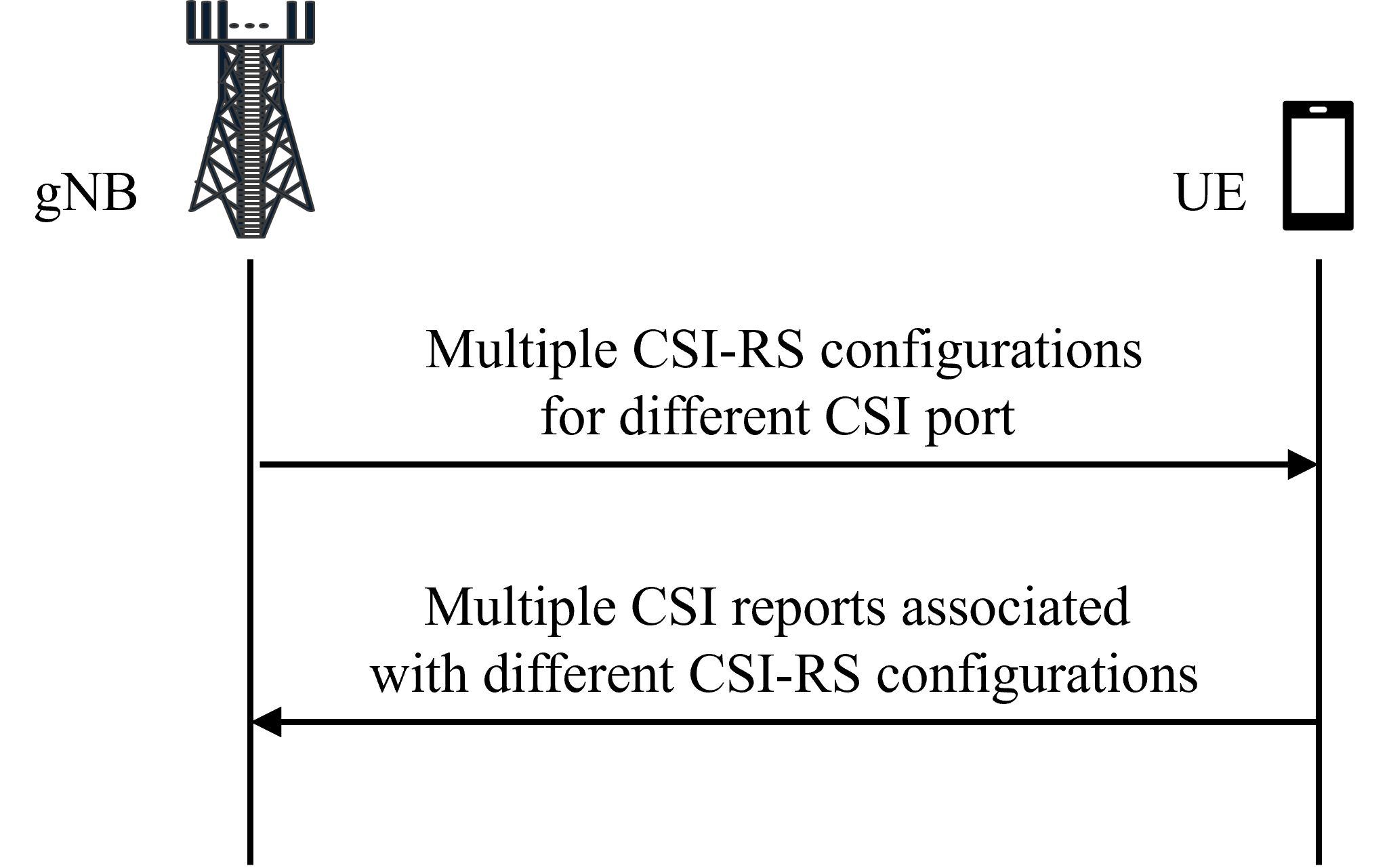}
    \caption{Illustration of multiple \acrshort{CSI-RS} framework}
    \label{fig:multicsi}
\end{figure}

We consider a network scenario where \gls{mMIMO} \gls{gNB} serves the set $\mathcal{N} \coloneqq \{1,2,\ldots,N\},~|\mathcal{N}| = N$ of active, multi-antenna \glspl{UE} in \gls{DL} on same time-frequency resources. The transmission opportunity in the time-domain is represented using a slot, which is denoted by $t \in \mathbb{N}^+$. In the frequency-domain, the total available bandwidth is divided into several \glspl{RB} denoted by the set $\mathcal{B} \coloneqq \{1,2,\ldots,B\},~|\mathcal{B}| = B$. We assume that \gls{gNB} assigns a subset $\mathcal{B}_n \subseteq \mathcal{B},~|\mathcal{B}_n| = B_n$ of \glspl{RB} to each \gls{UE} $n$ in an orthogonal manner. We also assume that \gls{gNB} configures each \gls{UE} with multiple \gls{CSI-RS} configurations following~\cite{3gpp.38.214} (see Fig.~\ref{fig:multicsi}). Firstly, each \gls{CSI-RS} configuration corresponds to \gls{gNB} operating with a predefined number of \gls{CSI-RS} ports, thereby facilitating antenna adaptation. Furthermore, the number of \glspl{RB} required for conveying a \gls{CSI-RS} configuration to a \gls{UE} is directly proportional to the number of \gls{CSI-RS} ports. Let $\mathcal{M} \coloneqq \{1,2,\ldots,M\},~|\mathcal{M}| = M$ denote the set of all \gls{CSI-RS} configurations. Here, each $m$ corresponds to \gls{gNB} operating with a predefined set of \gls{CSI-RS} ports and \glspl{TRX}. Furthermore, $m = M$ corresponds to the baseline scenario, where \gls{gNB} transmits in \gls{DL} with all available \glspl{TRX}. For other cases with $m < M, \forall~m \in \mathcal{M}/\{M\}$, we assume that there is a predefined mapping to determine which \gls{CSI-RS} ports and \glspl{TRX} to operate at \gls{gNB}. 

As a consequence of \glspl{UE} processing $M$ \gls{CSI-RS} configurations, \gls{gNB} receives $M$ number of \gls{CSI} reports from each \gls{UE} $n$. A \gls{CSI} report from \gls{UE} primarily comprises \gls{CQI}, and additionally a rank indicator for multi-stream communication and a precoder matrix indicator. In this work, we assume that the \gls{CQI} is available as a wideband metric, meaning, it is obtained at a \gls{UE} by averaging the estimated \gls{SINR} across the entire bandwidth $B$. Furthermore, \gls{gNB} assigns a \gls{MCS} value to each \gls{UE} $n$ based on the corresponding received \gls{SINR} estimates. Firstly, the set of \gls{MCS} values that can be assigned to \glspl{UE} is denoted by $\mathcal{K} = \{1,2,\ldots,K\},~|\mathcal{K}| = K$ w.l.o.g. Moreover, each \gls{MCS} $k \in \mathcal{K}$ is associated with a transmission rate $R(k)$. The link reliability of each \gls{UE} $n$ is evaluated using the achievable \gls{BLER}, which is denoted by $\epsilon_n(\gamma_{n,m};k)$. Moreover, the desired target \gls{BLER} for the first transmission attempt is denoted by $\Bar{\epsilon}$. Furthermore, the \gls{SINR} estimate at a \gls{UE} $n$ and associated with the $m^{\text{th}}$ \gls{CSI} configuration is denoted as $\gamma_{n,m} \coloneqq S_{n,m}\cdot\hat{\alpha}_{n,m}$. Here, $\hat{\alpha}_{n,m} \in \mathbb{R}$ collectively represents the channel and noise-plus-interference estimate of a \gls{UE} $n$ associated with the $m^{\text{th}}$ \gls{CSI} report. The term $S_{n,m}$ represents the \gls{PSD} of a \gls{UE} $n$ with respect to the $m^{\text{th}}$ \gls{CSI-RS} configuration and is given by: 
\begin{align}
\label{eq:psdexpression}
    S_{n,m} = \frac{P_{n,m}}{B_n},
\end{align}
where, $P_{n,m} \leq \Bar{P}_m$ is the power allocated to \gls{UE} $n$ when $m^{\text{th}}$ \gls{CSI-RS} configuration is used. The term $\Bar{P}_m$ is the maximum allowable transmit power associated with the $m^{\text{th}}$ \gls{CSI-RS} configuration. Finally, the number of bits pending for transmission at $n^{\text{th}}$ \gls{UE}'s buffer is denoted by $Q_n \leq \Bar{Q}$, with $\Bar{Q}$ being the maximum number of bits allowed in the buffer. Based on the notation presented here, an overview of \gls{MAC} operations are provided in the next section.

\section{MAC Layer Operations}
\label{sec:problemformulation}

In this section, we first discuss the resource scheduling aspects of the \gls{MAC} scheduler. This is followed by an overview of \gls{LA} schemes explained from the perspective of antenna adaptation and the \acrshort{POLITE} scheme. Firstly, given the set of active \glspl{UE} $\mathcal{N}$, the task of the \gls{MAC} layer is to select appropriate \gls{MCS} and allocate per-slot resources or \glspl{RB} among these \glspl{UE}. In practice, resource scheduling is performed using the widely adopted \gls{PF} scheduler due to its strong performance guarantees in \acrshort{MIMO} systems~\cite{sblee2009dlmimofdps, smandelli2021polite}. More specifically, the \gls{PF} scheduler allocates \glspl{RB} to \glspl{UE} based on a per-\gls{UE} metric obtained on each \gls{RB} $b$ as follows: 
\begin{align}\label{eq:pfmetric}
    X_n(b) = \frac{R_{n}(k;b)}{R_{n,\text{avg}}}, ~\forall~b \in \mathcal{B}.
\end{align}
Here, $R_{n}(k;b)$ is the \gls{MCS}-dependent instantaneous rate of \gls{UE} $n$ on the $b^{\text{th}}$ \gls{RB} and $R_{n,\text{avg}}$ is the running average of $n^{\text{th}}$ \gls{UE}'s throughput. In essence, the \gls{PF} metric allocates a \gls{RB} to a \gls{UE} with high spectral efficiency relative to its recent throughput history. Doing so, the \gls{PF} metric ensures that \glspl{RB} are proportionately allocated to \glspl{UE} on an average in time. 

As a consequence of considering the wideband \gls{CQI} in this work, Equation~\eqref{eq:pfmetric} reduces to a \textit{single metric} per-\gls{UE} across the entire bandwidth $B$. This also facilitates the use of greedy approaches to solve the originally NP-hard \gls{RB} allocation problem. One such approach involves allocating contiguous subset of \glspl{RB} to \glspl{UE} simply based on the estimated number of \glspl{RB}, which can be obtained as a function of the \gls{MCS} value~\cite{sblee2009dlmimofdps}. Since \gls{LA} is essentially responsible for determining the \gls{MCS} values of \glspl{UE}, the remainder of this section focuses on the \gls{LA} schemes considered in this work.

The \gls{LA} is performed for every \gls{UE} $n \in \mathcal{N}$ to determine the optimal \gls{MCS} by solving a rate maximization problem with \gls{BLER} requirements incorporated as constraints. Especially when each \gls{UE} $n \in \mathcal{N}$ is configured with $M$ \gls{CSI-RS} configurations, \gls{LA} is also performed for all $M$ configurations~\cite{3gpp.38.300}. In this case, the \gls{LA} problem at \gls{UE} $n \in \mathcal{N}$ for the $m^{\text{th}}$ \gls{CSI-RS} configuration can be written as follows:
\begin{subequations}
    \label{eq:multicsila}
    \begin{align}
        \textbf{P1:} \maximize_{k}~& R_{n}(k) \label{eq:p1obj1}\\
        \text{subject to }&\epsilon_n(\gamma_{n,m}; k) \leq \Bar{\epsilon} \label{eq:p1cons1} \\
        &S_{n,m} \leq \Bar{S}_{n,m},~k \in \mathcal{K},~\forall~m \in \mathcal{M}  \label{eq:p1cons2}.
\end{align}    
\end{subequations}
Here, the term $\Bar{S}_{n,m}$ is the maximum allowable \gls{PSD} derived with respect to $\Bar{P}_m$. Thus, the change in maximum allowable transmit power due to operating with multiple \gls{CSI-RS} configurations is reflected through the right hand side of the constraint~\eqref{eq:p1cons2}. Moreover, it can be observed that \textbf{P1} is a straightforward extension of baseline \gls{LA} with $m = M$ to the network setup with multiple \gls{CSI-RS} configurations. As a consequence of this, the solution approach of the baseline \gls{LA} can also be extended to any $m < M$, where the \gls{MCS} $k$ is greedily determined in the reference configuration using an upper bound on constraints~\eqref{eq:p1cons1} and~\eqref{eq:p1cons2}. 

\subsection{Overview of POLITE}
\label{sec:politela}
The original concept of \acrshort{POLITE}~\cite{smandelli2021polite} is essentially aimed at minimizing \gls{PSD} of \gls{DL} transmissions by refining \glspl{UE}' \gls{MCS} values determined using the baseline \gls{LA} with $m = M$. More specifically, \acrshort{POLITE} uses the buffer information of \glspl{UE} available at \gls{MAC} layer to effectively reduce the \gls{MCS} values of \glspl{UE} while still satisfying the target \gls{BLER} requirements. Consequently, this leads to an increase in the number of \glspl{RB} allocated to \glspl{UE}, thereby enabling \gls{DL} transmissions with an overall lower \gls{PSD} compared to the baseline \gls{LA}. Formally, \acrshort{POLITE} concept can be expressed in terms of \gls{LA} using the following optimization problem:
\begin{subequations}
    \label{eq:politela}
    \begin{align}
        \textbf{P2:} \minimize_{k^{\prime}}~& S_{n,m}(k^{\prime}) \label{eq:p2obj1}\\
        \text{subject to }& R_n(k^{\prime}) \geq \beta_n\cdot R_n(k) \label{eq:p2cons1} \\
        &\epsilon_n(\gamma_{n,m};k^{\prime}) \leq \Bar{\epsilon} \label{eq:p2cons2} \\
        &k^{\prime} \leq k,~\forall~k,k^{\prime} \in \mathcal{K},~\forall~n \in \mathcal{N}.
\end{align}    
\end{subequations} 
Firstly, it can be observed that \gls{PSD} $S_{n,m}$ is now a function of the \gls{MCS} value $k^{\prime}$. This dependency on \gls{MCS} is reflected in~\eqref{eq:psdexpression} through the denominator, rendering $B_n$ as a function of \gls{MCS} $k$ (or $k^{\prime}$ in~\eqref{eq:politela}). Secondly, the objective and the constraints are reversed in \textbf{P2} when compared to baseline \gls{LA}. Especially, the inclusion of constraint~\eqref{eq:p2cons1} ensures that the achievable rate is not affected as a consequence of operating with a lower \gls{MCS} due to \acrshort{POLITE}. Finally, the \gls{BLER} constraint in~\eqref{eq:p2cons2} is the same as the baseline \gls{LA}.

The solution approach of \acrshort{POLITE} revolves around modeling the linear rate-adaptation parameter $\beta_n$ in the constraint~\eqref{eq:p2cons1}. While there are several ways of evaluating the $\beta_n$ term~\cite{mandelli2022reducing}, in this work we resort to using the load-driven \acrshort{POLITE} variant~\cite{smandelli2021polite}. More specifically, the $\beta_n$ value is derived as a function of the buffer value $Q_n$ of a \gls{UE} $n$ and the total bandwidth $B$ at current slot as in~\cite{smandelli2021polite}. Accordingly, this provides a way to balance \gls{PSD} reduction without compromising on the achievable throughput. Formally, $\beta_n$ value of \gls{UE} $n$ following the load-driven approach can be expressed as follows:
\begin{align}\label{eq:politebeta}
    \beta_n \coloneqq \min\biggl[\chi\cdot \sum_{n \in \mathcal{N}}\frac{Q_n}{R_{n, \text{avg}}\cdot B}, 1\biggr],
\end{align}
where, $\chi \in (0,1]$ is a predefined multiplicative factor used to avoid abrupt changes in $\beta_n$ and $R_{n, \text{avg}}$ is the time-averaged throughput of \gls{UE} $n$ as described above. Firstly, evaluating per-\gls{UE} $\beta_n$ in this manner is termed load-driven \acrshort{POLITE}, since the users' packet arrivals influence the \gls{RB} usage or the network load of the current slot $t$. Another noticeable aspect of~\eqref{eq:politebeta} is that despite indexing with respect to each \gls{UE} $n$, $\beta_n$ is derived as a joint metric across all the associated \glspl{UE} in the network.

\section{Dynamic Antenna and Power Adaptation}
\label{sec:proposedsolution} 

In this work, a dynamic scheme for joint antenna and power adaptation is proposed, which opportunistically maximizes \gls{NES} at \gls{gNB} while inherently accounting for user throughput. More specifically, both antenna adaptation and power adaptation decisions are optimized together on a per-slot basis by leveraging the \gls{LA} schemes discussed in the previous section. In this regard, antenna adaptation is performed first as it results in a higher reduction in \gls{gNB} \gls{PC}~\cite{swesemann2023eemimo}. More specifically, number of active \glspl{TRX} are determined by solving the \gls{LA} problem in~\eqref{eq:multicsila}. This is followed by performing \gls{POLITE}-based power adaptation as described in~\eqref{eq:politela}. The motivation behind using \gls{POLITE} after antenna adaptation is to efficiently counter the intra-cell interference. More specifically, as \gls{gNB} operates with fewer \glspl{TRX}, the beams used for \gls{DL} transmissions to \glspl{UE} become wider, thereby causing interference in the frequency-domain. However, as \gls{POLITE} is more likely to reduce the \gls{MCS} of \glspl{UE} with higher \gls{MCS}, it effectively leads to \gls{DL} transmissions with reduced transmit power and eventually less interference. Thus, sequentially performing antenna adaptation and \gls{POLITE} through efficient \gls{LA} and \gls{RB} allocation leads to maximizing \gls{NES}, while delivering throughput to \glspl{UE}. A detailed overview of the steps involved in the proposed scheme are presented in \textbf{Alg.~\ref{alg:pseudocode}}.  

In \textbf{Alg.~\ref{alg:pseudocode}}, step~$1$ is evaluated using Equation~\eqref{eq:pfmetric} based on \gls{MCS} obtained using the reference configuration with $m=M$. Furthermore, steps~$2$~\textemdash~$10$ correspond to antenna adaptation, where the number of active antenna ports for \gls{DL} transmissions in the current slot are determined in a greedy manner. Here, $\mathcal{M}^*$ in step~$2$ refers to the set $\mathcal{M}$ arranged in the reverse order. More importantly, antenna adaptation is performed if and only if the corresponding \gls{RB} allocation allows for emptying the buffers of all the associated \glspl{UE}. Steps~$13$ and~$14$ correspond to \gls{POLITE} \gls{LA} as described in Section~\ref{sec:politela}, where $m = m^{\prime}$ is substituted instead of $m = M$ as in~\eqref{eq:politela}. In steps~$15$\textemdash~$19$, any remaining \glspl{RB} are distributed among \glspl{UE} whose buffers still have pending bits. It can be seen from \textbf{Alg.~\ref{alg:pseudocode}} that antenna and power adaptation decisions effectively take into account the \acrshort{CSI} and buffer information of \glspl{UE} at each slot. Secondly, \textbf{Alg.~\ref{alg:pseudocode}} essentially works by leveraging the \gls{PSD} expression in~\eqref{eq:psdexpression}. More specifically, the steps corresponding to antenna adaptation result in reduced transmit power $P_{n,m}$ in~\eqref{eq:psdexpression}. Finally, the steps associated with \gls{POLITE} \gls{LA} and the remaining steps result in an overall reduction of $S_{n,m}$ due to a decrease in \glspl{UE}' \gls{MCS} values and an increase in the allocated \glspl{RB}.
\begin{algorithm}[ht!]
    \caption{\strut Dynamic Antenna and Power Adaptation Scheme}
    \label{alg:pseudocode}
    \begin{algorithmic}[1]
    \INIT $m \gets M, m^{\prime} \gets m, \Tilde{k}_n \gets 0,~\forall~n \in \mathcal{N}$.
    \INPUT $k_n,~\forall~n \in \mathcal{N}$ using reference configuration $m=M$.
    \State Arrange \glspl{UE} in the descending order of $\{X_n\}_{n=1}^N$~\eqref{eq:pfmetric}. 
    \ForAll{$m \in \mathcal{M}^*/\{M\}$}
      \State $\Tilde{k}_n \gets k_n,~\forall n \in \mathcal{N}$. 
      \State Obtain $k_n$ and $B_n$ by solving~\eqref{eq:multicsila}, $\forall n \in \mathcal{N}$.
      \If{$Q_n,~\forall n \in \mathcal{N}$ can be emptied}
        \State $m^{\prime} \gets m$.
      \Else{}
      \State $m^{\prime} \gets m+1$.
      \State $k_n \gets \Tilde{k}_n,~\forall~n \in \mathcal{N}$.
      \State \Goto{alg:politestart}.
      \EndIf
    \EndFor
    \State Compute $\beta_n,~\forall~n \in \mathcal{N}$ following~\eqref{eq:politebeta}. \label{alg:politestart}
    \State Update $k_n$ and $B_n$ by solving~\eqref{eq:politela}, $\forall n \in \mathcal{N}$.
    \If{$B - \sum_{n \in \mathcal{N}}B_n > 0$}
      \State Recompute $\{X_n\}_{n=1}^N$ using~\eqref{eq:pfmetric}.
      \State Arrange \glspl{UE} in the descending order of $\{X_n\}_{n=1}^N$.
      \State Allocate remaining \glspl{RB} among \glspl{UE} with $Q_n > 0$.
    \EndIf
    \State \Return $m^{\prime}, k_n, B_n,~\forall~n \in \mathcal{N}$.
    \end{algorithmic}
\end{algorithm}

\section{Numerical Simulations}
\label{sec:numericalsimulations}

In this section, we present the performance evaluations of \textbf{Alg.~\ref{alg:pseudocode}} using numerical simulation results. To this extent we use a \gls{3GPP} compliant system level simulator modeled using the design principles presented in~\cite{kpedersen2024slstutorial}. However, prior to presenting the numerical results, we first briefly discuss the \glspl{KPI} and the simulation methodology used for the performance evaluation.

\subsection{Simulation Methodology}

In this work, we use two \glspl{KPI} to evaluate the performance of the algorithms considered for numerical simulation results. The first \gls{KPI} corresponds to \gls{UPT}, which is defined as the average application layer throughput per \gls{FTP3} packet transmission. Here, the averaging is performed both with respect to the number of transmitted packets of all \glspl{UE} and the total simulation duration. The second \gls{KPI} corresponds to \gls{NES} expressed in terms of \gls{gNB} \gls{PC}, which is evaluated using the \gls{3GPP} \gls{PC} model~\cite{3gpp2023studynes} as described below. 

The \gls{3GPP} \gls{PC} model is characterized by different sleep states, namely the micro, light and deep sleep states. Firstly, no transmission or reception takes place when \gls{gNB} operates in a sleep state. Secondly, each sleep state corresponds to the case where one or more hardware components of \gls{gNB} are switched off, with majority of components being switched off in the deep sleep state. Furthermore, the \gls{PC} associated with each sleep state is expressed in relative units in reference to the deep sleep state \gls{PC}. The use of relative units instead of absolute units (e.g. Watts) provides an efficient abstraction of \gls{gNB} implementations from different telecommunication network equipment vendors. 

The \acrshort{3GPP} \gls{PC} model also defines active state \gls{PC} values resulting from dynamic operations of \gls{gNB} in each slot. An active state here refers to the time period during which data transmission takes place in \gls{DL} and/or uplink directions. As this work pertains to \gls{DL} transmissions, only the relevant \gls{DL} \gls{PC} expressions are provided here. The \gls{DL} \gls{PC} in the active state due to per-slot \gls{gNB} operations are expressed as a combination of static and dynamic terms, which can be formally expressed as follows:
\begin{equation} \label{eq:dynamicpower}
    \text{PC}_{\text{active}} = s_a \left( s_f \cdot s_p\cdot \frac{P_{\text{dyn, joint}}}{\eta(s_f, s_p)} + P_{\text{dyn, ante}} \right) + P_{\text{static}}.
\end{equation}
The parameters $s_a$, $s_f$ and $s_p$ denote the ratio of active \glspl{TRX} to the total \glspl{TRX}, the ratio of occupied bandwidth to the system bandwidth, and the ratio of \gls{PSD} between the \gls{DL} transmission and reference configuration from~\cite[Table II]{laselva20246gnes}. Furthermore, $\frac{P_{\text{dyn, joint}}}{\eta(s_f, s_p)}$ is the \gls{PC} associated with the power amplifier, which is a function of the transmit power, number of \glspl{TRX} and the operating bandwidth. The term $\eta$ is the power amplifier efficiency parameter, which is modeled as a function of $s_f$ and $s_p$ factors as described in~\cite{laselva20246gnes}. Furthermore, $P_{\text{dyn, ante}}$ corresponds to \gls{PC} associated with hardware components such as digital front-end, which scales only as function of the active \glspl{TRX}. Finally, $P_{\text{static}}$ corresponds to \gls{PC} of essential hardware components which are required to keep \gls{gNB} operational in active and sleep states.

The results are generated for four different traffic or network load conditions, namely low, light, medium and high loads as specified in~\cite{3gpp2023studynes}. The load conditions are modeled as a function of packet arrivals at \glspl{UE}, where packets are generated using the \gls{FTP3} traffic model. Furthermore, we consider five baseline schemes in this work. Of these, three baselines correspond to static antenna configurations with $8$, $16$ and $32$ maximum number of \gls{gNB} \glspl{TRX}, which are referred to as \verb|Static8|, \verb|Static16| and \verb|Static32|, respectively. In addition to these baselines, we also consider a dynamic antenna adaptation and a power adaptation scheme as baselines in this work. In the antenna adaptation scheme, referred to as \verb|Antenna Adaptation| in the results, the number of active \glspl{TRX} are dynamically determined from the set $\{32,16,8\}$. Furthermore, \acrshort{POLITE} is used as the power adaptation baseline and is referred to as \verb|Power Adaptation| in the numerical results. Finally, the parameters used for generating the numerical results are summarized in Table~\ref{tab:numericalresults}.

\begin{table}[ht]
    \centering
    \begin{tabular}{c|c}
    \hline
    \textbf{Parameter} & \textbf{Value} \\
    \hline
    Number of \glspl{gNB} & $21$ \\
    Total number of \glspl{UE} & $210$ \\
    Maximum number of \gls{gNB} \glspl{TRX} & 32 \\
    \gls{UE} configuration & Single antenna \glspl{UE} \\
    Channel model & $38.901$ UMa NLoS~\cite{3gpp.38.901}\\ 
    \acrshort{MIMO} mode & Single User MIMO \\
    Duplexing protocol & TDD                  \\
    Bandwidth & 100 MHz                 \\ 
    \acrshort{OFDM} subcarrier spacing   & 30 kHz                  \\ 
    Carrier frequency & 3.5 GHz\\  
    Maximum transmit power at \gls{gNB} & 52 dBm \\
    Precoder type & Type-I\\
    MCS table & Table 5.1.3.1-2~\cite{3gpp.38.214} \\
    Packet arrival rate at \glspl{UE} & $\{100, 180, 340, 500\}$ \\
    Packet size & 5 Kbytes \\
    Maximum HARQ retransmissions & 4\\
    Number of simulation drops & $10$\\
    Number of slots per drop & $28000$ \\
    \hline
    \end{tabular}
    \caption{Numerical Simulation Parameters}
    \label{tab:numericalresults}
\end{table}

\subsection{Results Discussion}

\begin{figure}[ht!]%
    \centering
    \includegraphics[width=1\linewidth]{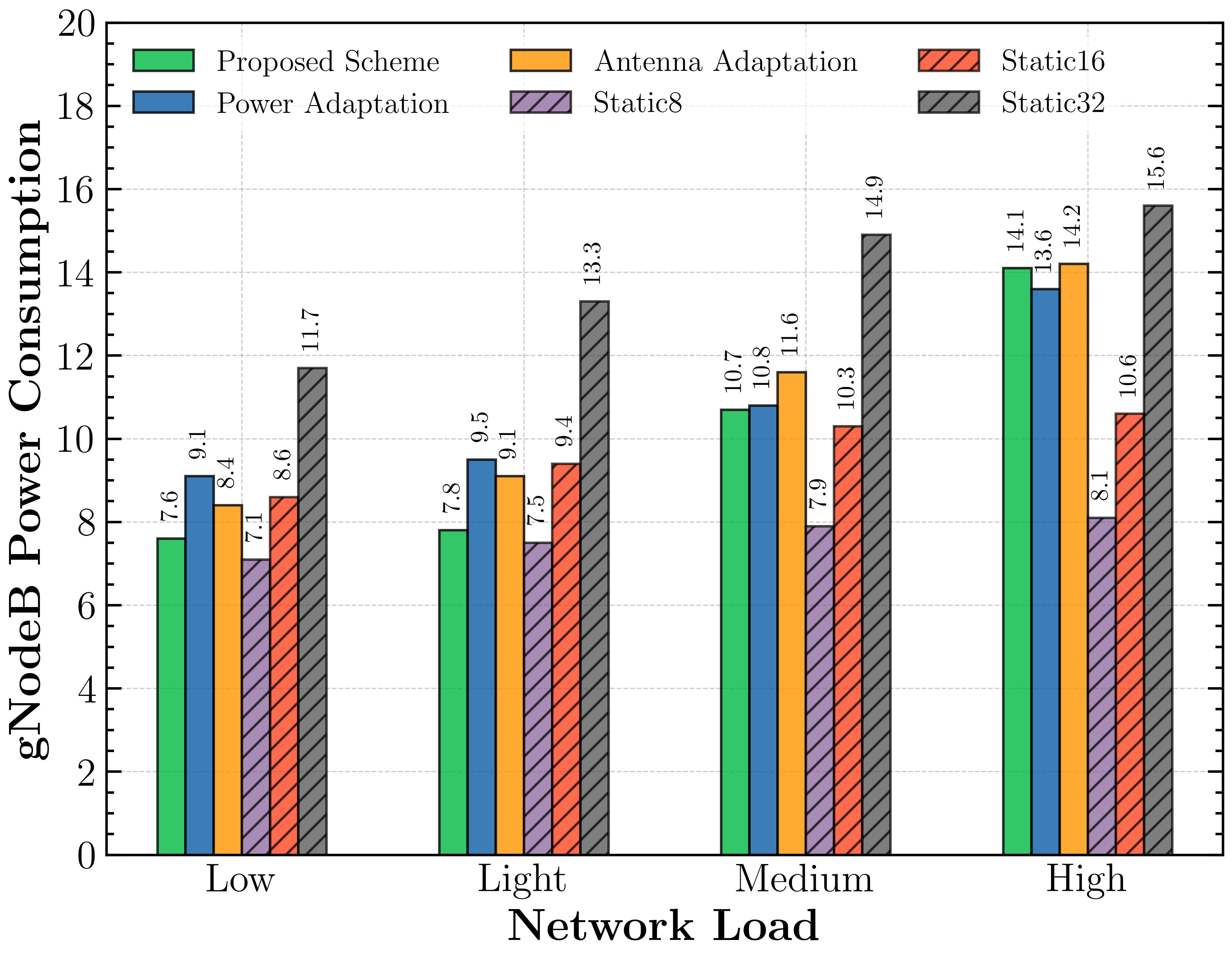}
    \caption{Comparison of power consumption vs network load}
    \label{fig:nesgaincomparison}
\end{figure}
\begin{figure}[ht!]
    \centering
    \includegraphics[width=1\linewidth]{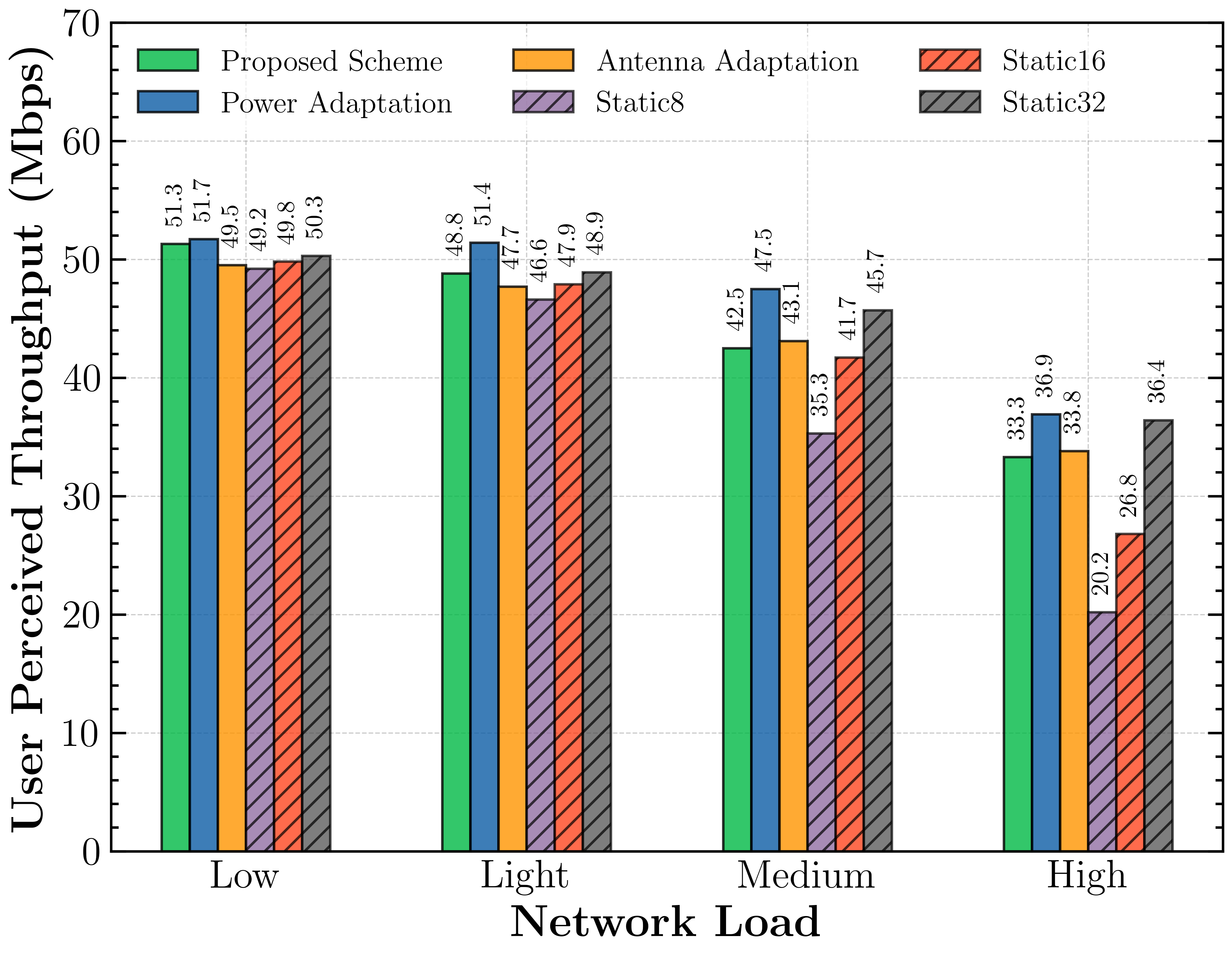}
    \caption{Comparison of user throughput vs network load}
    \label{fig:uptgaincomparison}
\end{figure}

We begin the performance evaluation by comparing \gls{gNB} \gls{PC} and \gls{UPT} performance in Figs.~\ref{fig:nesgaincomparison} and~\ref{fig:uptgaincomparison}, respectively. Firstly, the proposed scheme \textit{consistently} results in a lower \gls{PC} compared to \verb|Static32| in all load conditions. More importantly, \gls{NES} here is achieved without any significant degradation in \gls{UPT}. Secondly, \gls{PC} of the proposed scheme is on par with \verb|Static8| in low and light load conditions. In medium and high load conditions, \verb|Static8| and \verb|Static16| comparably result in a lower \gls{PC}. However, both these baselines fair poorly in terms of \gls{UPT} compared to the proposed scheme. This establishes the need for a dynamic \gls{NES} scheme that can adapt to traffic and \gls{CSI} of \glspl{UE} as proposed in this work. 

Furthermore, \gls{PC} of the proposed scheme is less than that of \verb|Antenna Adaptation| and \verb|Power Adaptation| in low and light load conditions and it is comparable with both these schemes in medium and high load conditions. That said, \gls{UPT} of the proposed scheme is consistently on par with both these schemes in all load conditions. This shows the importance of jointly optimizing antenna and power adaptation schemes through efficient resource allocation, as described in \textbf{Alg.~\ref{alg:pseudocode}}. Overall, the proposed scheme strikes a good balance between \gls{NES} and \gls{UPT} under varying traffic and channel conditions.

\begin{figure}[ht!]%
    \centering
    \includegraphics[width=1\linewidth]{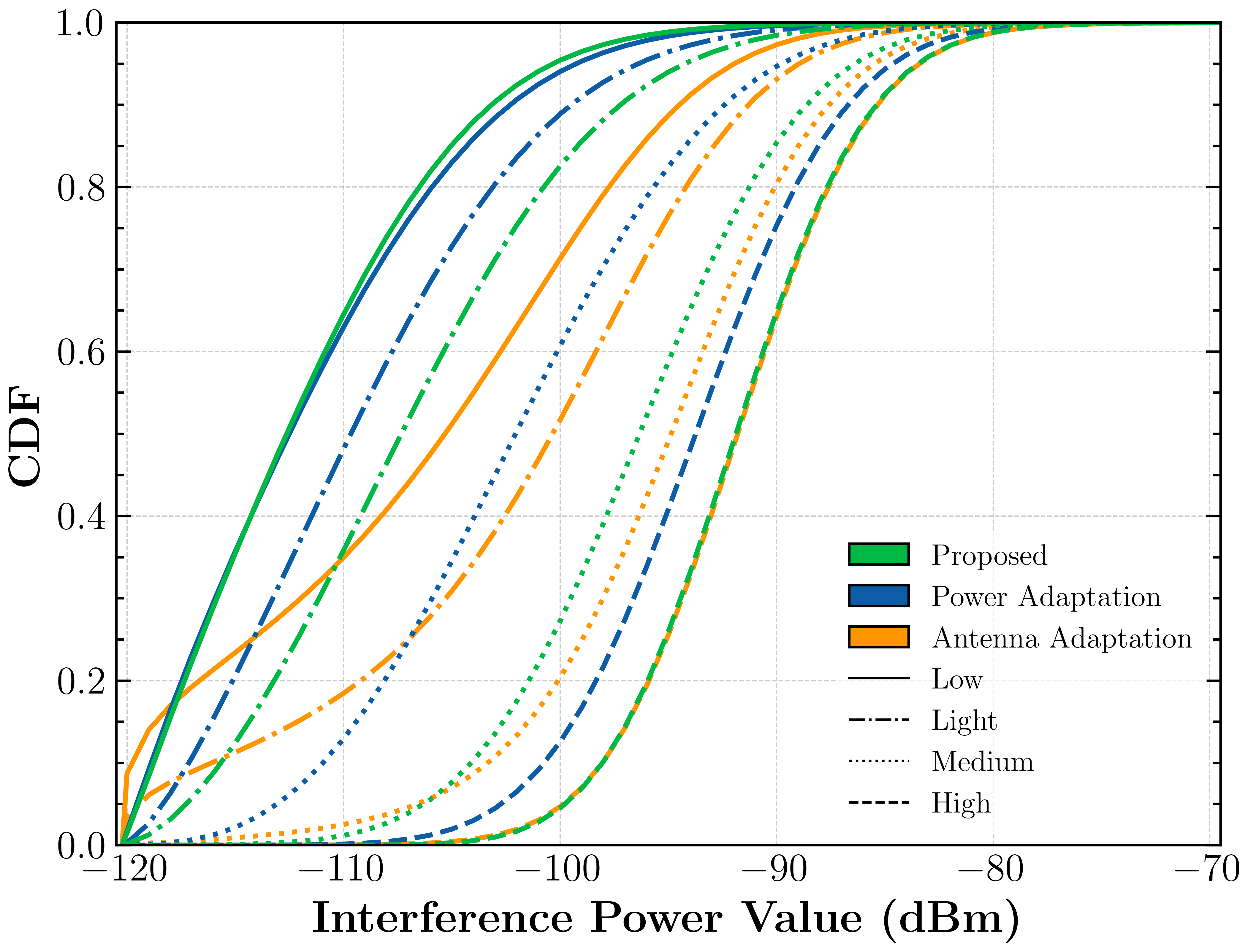}
    \caption{Empirical distribution of interference power values}
    \label{fig:Interference}
\end{figure}
\begin{figure}
    \centering
    \includegraphics[width=1\linewidth]{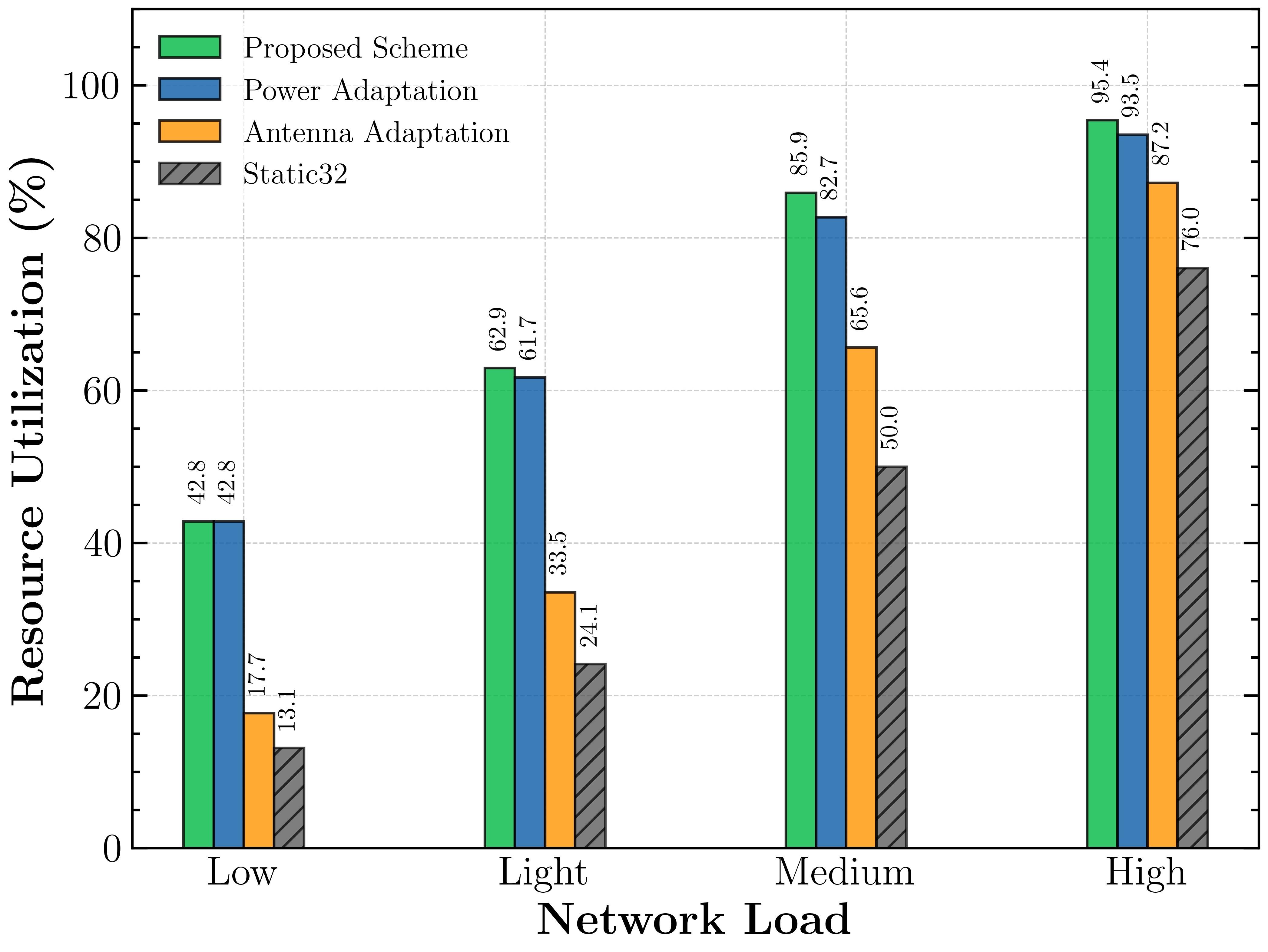}
    \caption{Comparison of resource utilization vs network load}
    \label{fig:PRButilization}
\end{figure}

In Fig.~\ref{fig:Interference}, the empirical distribution of intra-cell \gls{IPV} is plotted, where \gls{IPV} is sampled per \gls{RB}. Firstly, the proposed scheme induces very low intra-cell interference in low and light load conditions. While \gls{IPV} increases with the load, it is still induces less interference or is comparable with \gls{IPV} distribution of \verb|Antenna Adaptation|. This can be attributed to the interplay between \gls{LA} and resource allocation in the proposed scheme in \textbf{Alg.~\ref{alg:pseudocode}}. Finally, in Fig.~\ref{fig:PRButilization}, we compare the \gls{RB} utilization of the proposed scheme with the baselines. It can be seen that the proposed scheme consistently results in high \gls{RB} usage. Secondly, the \gls{RB} usage of the proposed scheme is almost comparable with that of the \gls{POLITE}-based power adaptation. This, together with \gls{NES} and \gls{UPT} performance in Figs.~\ref{fig:nesgaincomparison} and~\ref{fig:uptgaincomparison} shows that efficient resource allocation plays a key role in balancing \gls{UPT} and \gls{NES}. 

\section{Conclusions}
\label{sec:conclusions}

In this work, a dynamic scheme for joint antenna and power adaptation scheme is proposed at the \gls{MAC} scheduler to efficiently balance between \gls{NES} and \gls{UPT} in \gls{DL} at a \acrshort{mMIMO} \gls{gNB}. More specifically, the proposed scheme efficiently adapts to both \acrshort{CSI} and traffic conditions of the associated \glspl{UE}, which can be extremely beneficial in power-hungry \acrshort{mMIMO} deployments. Numerical simulation results generated using a \gls{3GPP} compliant simulator demonstrate that the proposed scheme achieves the intended balance between \gls{NES} and \gls{UPT} for all traffic load conditions. Especially in low and light load conditions, the proposed scheme achieves \gls{NES} gains of $\approx35\% - 41\%$ compared to static antenna adaptation baselines. This is noteworthy since antenna adaptation and \gls{POLITE}-based power adaptation are performed in a sequential manner by efficiently leveraging \gls{LA} and resource allocation. Future work will focus on solutions to overcome the \acrshort{CSI} acquisition limitations of the multiple \gls{CSI-RS} framework to further improve \gls{NES} and \gls{UPT} in \acrshort{6G} \gls{mMIMO}-based \glspl{gNB}.

\bibliographystyle{IEEEtran}
\bibliography{references}

\begin{thebibliography}{10}
\providecommand{\url}[1]{#1}
\csname url@samestyle\endcsname
\providecommand{\newblock}{\relax}
\providecommand{\bibinfo}[2]{#2}
\providecommand{\BIBentrySTDinterwordspacing}{\spaceskip=0pt\relax}
\providecommand{\BIBentryALTinterwordstretchfactor}{4}
\providecommand{\BIBentryALTinterwordspacing}{\spaceskip=\fontdimen2\font plus
\BIBentryALTinterwordstretchfactor\fontdimen3\font minus \fontdimen4\font\relax}
\providecommand{\BIBforeignlanguage}[2]{{%
\expandafter\ifx\csname l@#1\endcsname\relax
\typeout{** WARNING: IEEEtran.bst: No hyphenation pattern has been}%
\typeout{** loaded for the language `#1'. Using the pattern for}%
\typeout{** the default language instead.}%
\else
\language=\csname l@#1\endcsname
\fi
#2}}
\providecommand{\BIBdecl}{\relax}
\BIBdecl

\bibitem{muusitalo2025towards6g}
M.~A. Uusitalo, H.~Farhadi, M.~Boldi, M.~Ericson, G.~Fettweis, B.~M. Khorsandi, I.~L. Pavón, M.~Mueck, A.~Nimr, A.~Pärssinen, B.~Richerzhagen, P.~Rugeland, D.~Sabella, and H.~Wymeersch, ``{Toward 6G — Hexa-X Project's Key Findings},'' \emph{IEEE Communications Magazine}, vol.~63, no.~5, pp. 142--148, 2025.

\bibitem{zwang2024tutorialemimo}
Z.~Wang, J.~Zhang, H.~Du, D.~Niyato, S.~Cui, B.~Ai, M.~Debbah, K.~B. Letaief, and H.~V. Poor, ``A tutorial on extremely large-scale mimo for 6g: Fundamentals, signal processing, and applications,'' \emph{IEEE Communications Surveys \& Tutorials}, vol.~26, no.~3, pp. 1560--1605, 2024.

\bibitem{bjornson2017massive}
E.~Bj{\"o}rnson, J.~Hoydis, L.~Sanguinetti \emph{et~al.}, ``{Massive MIMO networks: Spectral, energy, and hardware efficiency},'' \emph{Foundations and Trends{\textregistered} in Signal Processing}, vol.~11, no. 3-4, pp. 154--655, 2017.

\bibitem{swesemann2023eemimo}
S.~Wesemann, J.~Du, and H.~Viswanathan, ``{Energy Efficient Extreme MIMO: Design Goals and Directions},'' \emph{IEEE Communications Magazine}, vol.~61, no.~10, pp. 132--138, 2023.

\bibitem{3gpp2023studynes}
3GPP, ``{Study on Network Energy Savings for NR (Release 18)},'' {3rd Generation Partnership Project (3GPP)}, Technical Report 38.864, 2023.

\bibitem{dlperez2022survey5gee}
D.~López-Pérez, A.~De~Domenico, N.~Piovesan, G.~Xinli, H.~Bao, S.~Qitao, and M.~Debbah, ``{A Survey on 5G Radio Access Network Energy Efficiency: Massive MIMO, Lean Carrier Design, Sleep Modes, and Machine Learning},'' \emph{IEEE Communications Surveys \& Tutorials}, vol.~24, no.~1, pp. 653--697, 2022.

\bibitem{dlaselva2024pathto6G}
D.~Laselva \emph{et~al.}, ``{White paper: The Path to {6G} with Unparalleled Energy Savings -- A {3GPP} Standardization Perspective},'' Nokia, Tech. Rep., 2024.

\bibitem{smandelli2021polite}
S.~Mandelli, A.~Lieto, P.~Baracca, A.~Weber, and T.~Wild, ``{Power Optimization for Low Interference and Throughput Enhancement for {5G} and {6G} systems},'' in \emph{2021 IEEE Wireless Communications and Networking Conference Workshops (WCNCW)}, 2021, pp. 1--7.

\bibitem{mmahossain2018eemmimoload}
M.~M.~A. Hossain, C.~Cavdar, E.~Björnson, and R.~Jäntti, ``{Energy Saving Game for Massive MIMO: Coping With Daily Load Variation},'' \emph{IEEE Transactions on Vehicular Technology}, vol.~67, no.~3, pp. 2301--2313, 2018.

\bibitem{nrajapaksha2024minimizing}
N.~Rajapaksha, J.~Mohammadi, S.~Wesemann, T.~Wild, and N.~Rajatheva, ``{Minimizing Energy Consumption in MU-MIMO via Antenna Muting by Neural Networks With Asymmetric Loss},'' \emph{IEEE Transactions on Vehicular Technology}, vol.~73, no.~5, pp. 6600--6613, 2024.

\bibitem{wparmudito2020loadawareee}
W.~Pramudito, E.~Alsusa, A.~Al-Dweik, K.~A. Hamdi, D.~K.~C. So, and T.~L. Marzetta, ``{Load-Aware Energy Efficient Adaptive Large Scale Antenna System},'' \emph{IEEE Access}, vol.~8, pp. 82\,592--82\,606, 2020.

\bibitem{peschiera2025optimizingtimespacepowerdomain}
\BIBentryALTinterwordspacing
E.~Peschiera, Y.~Agram, F.~Quitin, L.~V. der Perre, and F.~Rottenberg, ``{On Optimizing Time-, Space- and Power-Domain Energy-Saving Techniques for Sub-6 GHz Base Stations},'' 2025. [Online]. Available: \url{https://arxiv.org/abs/2505.15445}
\BIBentrySTDinterwordspacing

\bibitem{3gpp.38.214}
\BIBentryALTinterwordspacing
3GPP, ``{NR; Physical layer procedures for data},'' {3rd Generation Partnership Project (3GPP)}, Technical Specification (TS) 38.214, 07 2025, version 18.7.0. [Online]. Available: \url{https://portal.3gpp.org/desktopmodules/Specifications/SpecificationDetails.aspx?specificationId=3216}
\BIBentrySTDinterwordspacing

\bibitem{sblee2009dlmimofdps}
S.-B. Lee, S.~Choudhury, A.~Khoshnevis, S.~Xu, and S.~Lu, ``{Downlink MIMO with Frequency-Domain Packet Scheduling for 3GPP LTE},'' in \emph{IEEE INFOCOM 2009}, 2009, pp. 1269--1277.

\bibitem{3gpp.38.300}
\BIBentryALTinterwordspacing
3GPP, ``{NR; Physical layer procedures for data},'' {3rd Generation Partnership Project (3GPP)}, Technical Specification (TS) 38.300, 10 2025, version 18.7.0. [Online]. Available: \url{https://portal.3gpp.org/desktopmodules/Specifications/SpecificationDetails.aspx?specificationId=3216}
\BIBentrySTDinterwordspacing

\bibitem{mandelli2022reducing}
S.~Mandelli, A.~Lieto, M.~Razenberg, A.~Weber, and T.~Wild, ``{Reducing interference via link adaptation in delay-critical wireless networks},'' \emph{EURASIP Journal on Wireless Communications and Networking}, vol. 2022, no.~1, p. 109, 2022.

\bibitem{kpedersen2024slstutorial}
K.~I. Pedersen, R.~Maldonado, G.~Pocovi, E.~Juan, M.~Lauridsen, I.~Z. Kovács, M.~Brix, and J.~Wigard, ``{A Tutorial on Radio System-Level Simulations With Emphasis on 3GPP 5G-Advanced and Beyond},'' \emph{IEEE Communications Surveys \& Tutorials}, vol.~26, no.~4, pp. 2290--2325, 2024.

\bibitem{laselva20246gnes}
D.~Laselva, S.~Hakimi, M.~Lauridsen, B.~Khan, D.~Kumar, and P.~Mogensen, ``{On the Potential of Radio Adaptations for 6G Network Energy Saving},'' in \emph{2024 Joint European Conference on Networks and Communications \& 6G Summit (EuCNC/6G Summit)}, 2024, pp. 1157--1162.

\bibitem{3gpp.38.901}
\BIBentryALTinterwordspacing
3GPP, ``{Study on channel model for frequencies from 0.5 to 100 GHz},'' {3rd Generation Partnership Project (3GPP)}, Technical Specification (TS) 38.901, 05 2024, version 18.0.0. [Online]. Available: \url{https://portal.3gpp.org/desktopmodules/Specifications/SpecificationDetails.aspx?specificationId=3173}
\BIBentrySTDinterwordspacing

\end{thebibliography}

\end{document}